\documentclass[11pt,preprint,prd,nofootinbib]{revtex4}

\usepackage[pdftex]{graphicx}
\usepackage{latexsym,amsmath,amssymb,lmodern,float,url}
\usepackage{natbib}
\usepackage[pdftex,bookmarks,linktocpage,pdfpagelabels,plainpages=false,hyperfigures,linkcolor=blue,citecolor=blue,urlcolor=blue]{hyperref} 
\usepackage{xspace}
\usepackage{courier}
\hypersetup{colorlinks=true}

\usepackage{graphicx}
\usepackage{amsmath}
\usepackage{amssymb}

\graphicspath{{./figs/}}

\newcommand\be{\begin{equation}}
\newcommand\ee{\end{equation}}
\newcommand\bea{\begin{eqnarray}}
\newcommand\eea{\end{eqnarray}}
\newcommand{\appropto}{\mathrel{\vcenter{
  \offinterlineskip\halign{\hfil$##$\cr
    \propto\cr\noalign{\kern2pt}\sim\cr\noalign{\kern-2pt}}}}}

\begin{document}
\preprint{MI-TH-185}


\title{Catching p via s wave with  indirect detection}

\author{Steven~J.~Clark$^{\bf a}$}
\author{James~B.~Dent$^{\bf b}$}
\author{Bhaskar~Dutta$^{\bf a}$}
\author{Louis~E.~Strigari$^{\bf a}$}

\affiliation{$^{\bf a}$ Mitchell Institute for Fundamental Physics and Astronomy,
   Department of Physics and Astronomy, Texas A\&M University, College Station, TX 77845, USA}
\affiliation{$^{\bf b}$ Department of Physics, Sam Houston State University, Huntsville, TX 77341, USA}

\begin{abstract}
For many dark matter models, the annihilation cross section to two-body final states is difficult to probe with current experiments because the dominant annihilation channel is velocity or helicity suppressed. The inclusion of gauge boson radiation for three-body final states can lift the helicity suppression, allowing a velocity-independent cross section to dominate the annihilation process, and providing an avenue to constrain these models. Here we examine experimental constraints on dark matter that annihilates to two leptons plus a bremsstrahlung boson, $\bar{\ell}+\ell+\gamma/W/Z$. We consider experimental constraints on photon final states from Fermi-LAT using both diffuse photon data and data from dwarf spheroidal galaxies, and compare to the implied constraints from 21 cm measurements. Diffuse photon line searches are generally the strongest over the entire mass regime. We in particular highlight the model in which dark matter annihilates to two neutrinos and a photon, and show that these models are more strongly constrained through photon measurements than through existing neutrino bounds. 
\end{abstract}

\maketitle

\section{Introduction}

\par The annihilation cross section, $(\sigma v)_{\rm{ann}}$, is one of the key quantities that describes the nature of dark matter interactions with the Standard Model. The annihilation cross section in the early universe sets the relic abundance for thermally produced dark matter. There are several observational bounds on the annihilation cross section, for example from high energy gamma-ray data~\cite{Ackermann:2015zua,Archambault:2017wyh}, and from the cosmic microwave background (CMB)~\cite{Aghanim:2018eyx}. The redshifted 21 cm line arising from a spin-flip transition in neutral hydrogen gas prior to the era of recombination~\cite{Madau:1996cs} has recently been recognized as an important probe of dark matter annihilation and decay. The 21 cm measurements are sensitive to the annihilation cross section at redshifts $z \lesssim 15$~\cite{DAmico:2018sxd}.

\par The dark matter annihilation cross section can be expressed as a partial-wave expansion in powers of the square of the relative velocity between the annihilating particles~\cite{Landau:1959}. The evolving nature of the dark matter velocity implies that the leading order annihilation process may differ over the course of the evolution of the universe and during cosmological structure formation. For example, dark matter that is a Majorana fermion can naturally annihilate dominantly as a \emph{p}-wave process, $(\sigma v)_{\rm{ann}} \propto v^2$, during the era of dark matter freeze-out where the relative velocity squared is $v^2 \sim T/m_{\rm DM} \sim 1/20$. There exist $s$-wave annihilation channels to two fermion final states, for example, but these are chirality suppressed by mass ratio factors of $(m_f/m_{\rm DM})^2$~\cite{Goldberg:1983nd}. This can be contrasted with dark matter in the Galactic halo, which has a virial velocity $v^2 \sim 10^{-6}$, and thereby reduces the observational importance of the $p$-wave process. This implies that annihilation to three-body final states, which proceed via \emph{s}-wave annihilation due to the bremsstrahlung of a bosonic state~\cite{Bell:2011if}, can provide the leading order annihilation channel. 

\par Typically, three-body annihilation dominates over two-body annihilation as the mass of the particle mediating the interaction approaches that of the dark matter~\cite{Bergstrom:1989jr,Flores:1989ru,Baltz:2002we,Bringmann:2007nk,Bergstrom:2008gr,Barger:2009xe}. This also provides an alternative means of accessing the parameter space in the so-called compressed mass region, where the masses are nearly degenerate. However, the degeneracy does not need to be extreme in order to obtain a large effect from vector bremsstrahlung, which provides a natural region of parameter space that can be probed by experiment. 

\par These considerations suggest that, in the case of dark matter annihilation to fermion pairs, final state radiation and internal vector bremsstrahlung of photons and the electroweak $W/Z$ bosons are irreducible processes which present an interesting target for observational searches. Photon bremsstrahlung can produce a line-like spectral feature~\cite{Bell:2011if}, and the subsequent decay of the radiated $W$ and $Z$ bosons produce additional diffuse photon signals which provide a complementary avenue of investigation. 

\par In this work we examine constraints on dark matter annihilating predominantly through $s$-wave channels in the present universe as a result of electromagnetic and electroweak bremsstrahlung, but whose relic abundance is set in the early universe by $p$-wave annihilations. We consider Fermi-LAT observations of dwarf spheroidals (dSphs) and diffuse gamma-ray data. We are thus able to simultaneously probe both of these partial wave components for a single dark matter model. We demonstrate that current observations are able to constrain such models, which can be contrasted with the case of dark matter annihilating only through $p$-wave processes which is wholly inaccessible to observational limits. We discuss the constraints in the context of both thermal and non-thermal models. Bounds on the dark matter annihilation cross section from 21 cm observations are also presented along with a comparison to the constraints from CMB data.

\par We highlight in particular on the final state consisting of neutrinos only. This two-body final state is quite difficult to probe observationally, but the addition of a final state photon allows for more strict experimental constraints. A complication arises due to $SU(2)_L$ invariance, which makes producing a neutrinos-only final state as the dominant annihilation channel a non-trivial task, as one would expect the annihilation to also produce charged leptons. However we introduce a model which produces a neutrino-only final states without allowing final state charged lepton production, while respecting gauge invariance.

\par The remainder of this work is as follows. In Sec.~\ref{sec:brem} we review bremsstrahlung as a mechanism for lifting velocity suppression, and present annihilation spectra for final states of interest. In Sec.~\ref{sec:21cm} we discuss the role that the measurement of the 21 cm line can play in determining the strength of dark matter annihilations. Sec.~\ref{sct:results} discusses the signal targets and bounds derived from the Fermi satellite, along with the results from recent 21 cm observations of the EDGES collaboration. We conclude with a summary of these results in Sec.~\ref{sec:summary}.

\section{Lifting velocity suppression via bremsstrahlung}
\label{sec:brem}

\par In what follows we will adopt a single-component SUSY-inspired simplified model of dark matter consisting of a Majorana dark matter particle, $\chi$, whose fractional abundance gives the totality of the dark matter ($f_{\rm{DM}} = 1$), and which annihilates to Standard Model particles through $t-$ and $u-$ channel exchange of a colored scalar. 

\par The annihilation cross section for a pair of non-relativistic dark matter particles of total orbital angular momentum $L$ and relative velocity, $v$, is expressed as a partial wave expansion $\sigma v \propto v^{2L}$~\cite{Landau:1959}. Using general considerations (see for example~\cite{Bell:2008ey,Kumar:2013iva}), one finds that models with Majorana pair annihilation may naturally proceed dominantly through a \emph{p}-wave process, as it has \emph{s}-channel annihilations only through pseudoscalar, scalar, or axial-vector exchange, with an $L = 0$, \emph{s}-wave process arising only in models with pseudoscalar mediators (though the pseudoscalar typically couples through a Yukawa-like interaction, introducing a mass suppression in the same mold as chirality suppression). As is well-known~\cite{Goldberg:1983nd}, the axial-vector exchange also contributes an $L = 0$ partial-wave which is chirality suppressed for annihilation to light final state fermions of mass $m_f$ by the factor $m_f^2/m_{\rm DM}^2$. There also exist \emph{t}- and \emph{u}-channel annihilation modes through scalar exchange (as in SUSY and SUSY-inspired models) that produce chirality suppression, as can be seen through a Fierz transformation to the \emph{s}-channel where the axial-vector contribution is apparent~\cite{Bell:2010ei}.

\par Chirality suppression for annihilation to a pair of final state fermions may be evaded by annihilation to a three-body final state through the bremsstrahlung emission of a boson. This has been demonstrated in photon, gluon, and electroweak bremsstrahlung for $t$- and $u$-channel annihilation, as well as Higgstrahlung from an $s$-channel annihilation mode~\cite{Bergstrom:1989jr,Flores:1989ru,Baltz:2002we,Bringmann:2007nk,Ciafaloni:2011sa,Bell:2011if,Kumar:2016mrq}. For Galactic dark matter with virial velocities of $v \sim \mathcal{O}(10^{-3})$, there are regions of parameter space where the three-body final state process can dominate over the two-body final state. Specifically, the three-body final state process will increase relative to the two-body final state as the mediator mass approaches the dark matter mass. However, as we will demonstrate, the splitting between the dark matter mass and mediator mass does not need to be extremely fine tuned in order for a non-negligible effect to arise. This allows for the intriguing situation where the relic abundance of Majorana dark matter can be set by \emph{p}-wave annihilation while signals at later times are most strongly constrained by \emph{s}-wave processes induced through bremstrahlung. It should be emphasized that bounds on dark matter models must necessarily include the effects of such irreducible brem processes.

\par
\begin{figure*}[ht]
\begin{tabular}{cc}
\includegraphics[height=5cm]{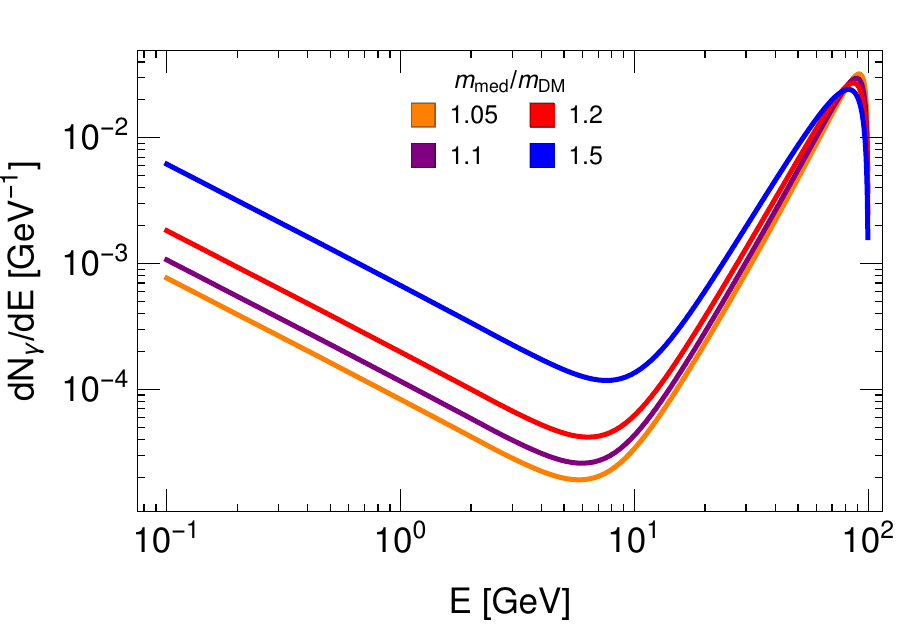} &
\includegraphics[height=5cm]{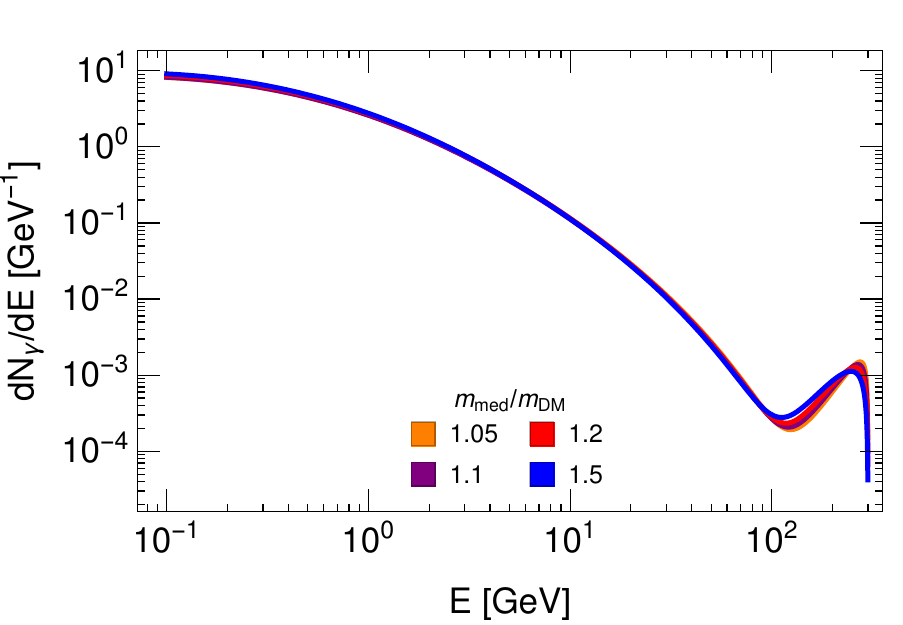}\\
\includegraphics[height=5cm]{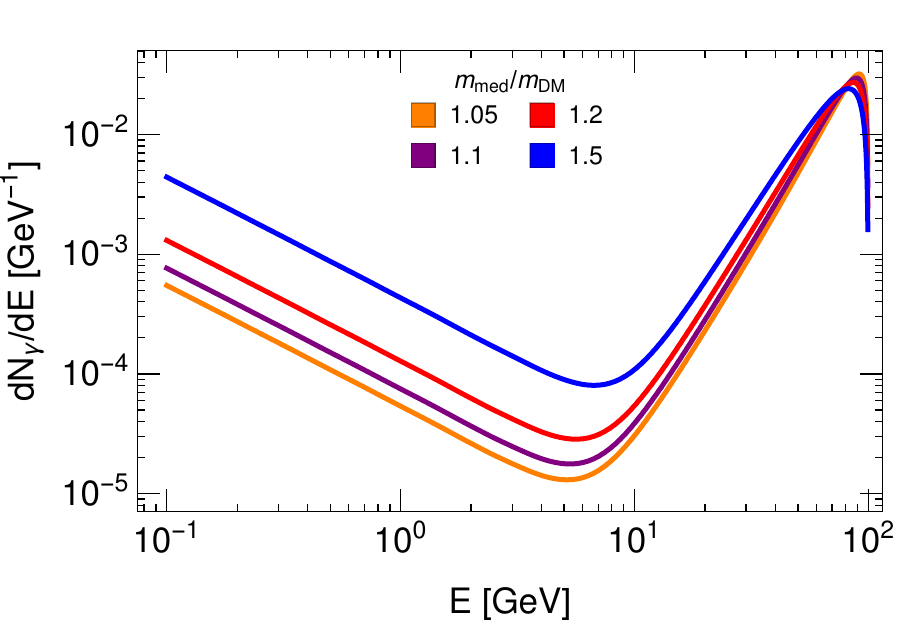} &
\includegraphics[height=5cm]{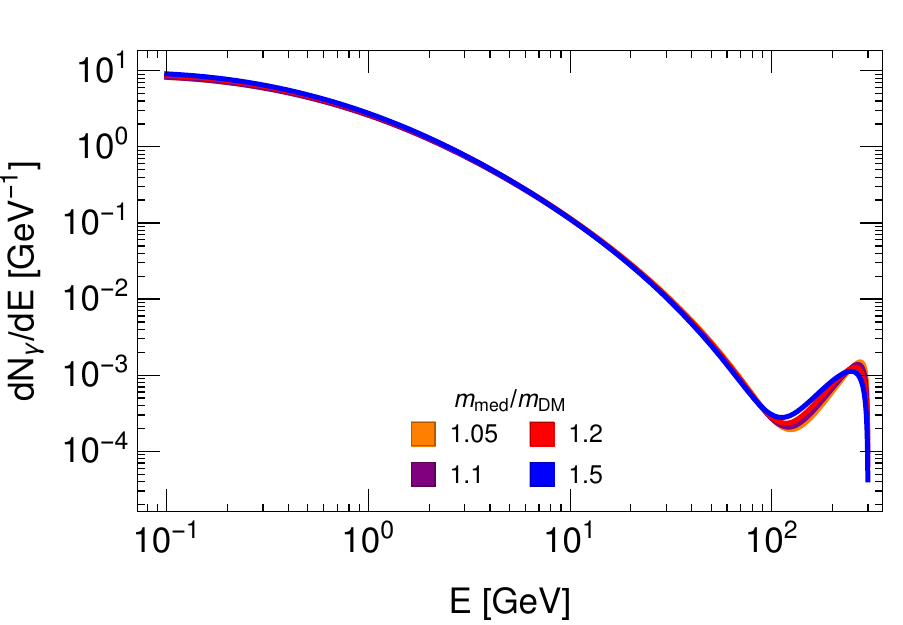}\\
\includegraphics[height=5cm]{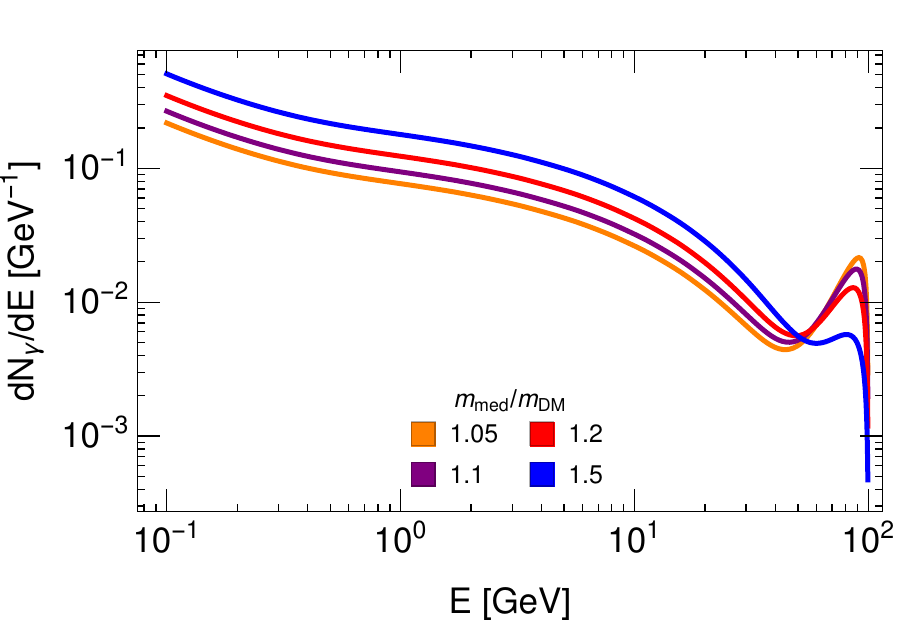} &
\includegraphics[height=5cm]{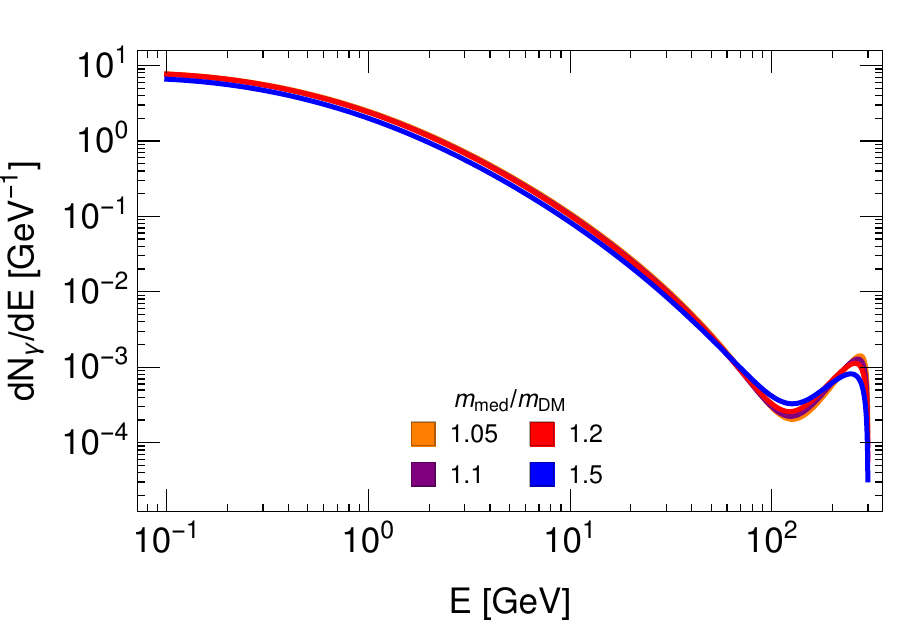}
\end{tabular}
\caption{Gamma-ray spectrum from a dark matter annihilation of $m_{\rm DM} = 100$ GeV (left) and $m_{\rm DM} = 300$ GeV (right) for various mediator mass ratios, $m_{\rm med}/m_{\rm DM}$. Lepton final states are electron (top), muon (middle), and tau (bottom). Only the two to two and photon bremsstrahlung are considered for the 100 GeV cases while 300 GeV includes the two to two and photon/W/Z bremsstrahlung.}
\label{fig:spectraCompare}
\end{figure*}

\begin{figure*}[ht]
\includegraphics[height=5cm]{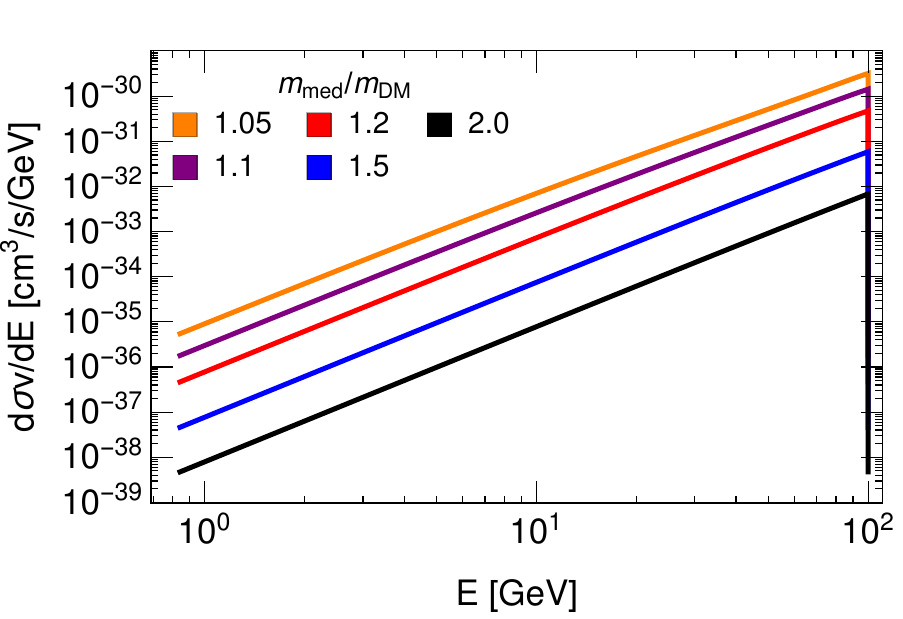}
\caption{Gamma-ray thermally averaged differential cross-section from a dark matter annihilation of $m_{\rm DM} = 100$ GeV into neutrino and gamma-ray final states for various mediator mass ratios, $m_{\rm med}/m_{\rm DM}$.}
\label{fig:spectraNuNu}
\end{figure*}

 Inclusion of the photon brem process induces a spectral feature that provides a target in line searches for gamma-ray observatories. This can be seen in the left column of Fig.~\ref{fig:spectraCompare}, which displays the photon spectrum for 100 GeV dark matter annihilating to charged fermion pairs, $e^+e^-$, $\mu^+\mu^-$, and $\tau^+\tau^-$ for different values of the mediator to dark matter mass ratio from 1.05 to 1.5. We classify these annihilations as two to two interactions. We also show the thermally averaged differential cross section for $\nu\nu$ cases in Fig.~\ref{fig:spectraNuNu} for mass ratios 1.05 to 2. We use the MSSM with a slepton mediator to calculate the $e,\,\mu,\,\tau$ final states, and introduce a new model to calculate the $\nu$ final states, described in the following paragraph. However, this analysis can be applied to any model with the same final states. This spectrum demonstrates the line-like feature that arises at the kinematic endpoint of the annihilation process, as well as showing that the lower energy spectral feature increases as the mediator mass approaches the dark matter mass. If the dark matter mass is large enough to produce on-shell $W/Z$ bosons, as in the right column of Fig.~\ref{fig:spectraCompare}, where the dark matter mass is 300 GeV, the line-like feature persists, but the low energy spectrum is enhanced from the $W/Z$ decays. 
 
\par In addition to searches with final state charged leptons, we also consider neutrino-only final states. Generally speaking, if a model contains a $\nu\bar{\nu}$ final state from DM annihilation, indirect detection becomes challenging (though bounds on the annihilation cross section can be determined~\cite{Beacom:2006tt,Albert:2016emp,Aartsen:2017ulx,ElAisati:2017ppn}). However,  the situation is improved if a $\nu\bar{\nu}\gamma$ final state is available. One can think about a possible model (for other models that produce a $\nu\bar{\nu}$ final state, see for example~\cite{ElAisati:2017ppn}) for such a scenario with the following Lagrangian:
\begin{equation}
{\cal{L}}\supset \lambda \phi \phi^\ast\rho\rho^\dagger+\lambda^\prime \bar{L}_v\rho\nu_R
\end{equation} 
Here $\rho=\begin{pmatrix}\rho^+\\\rho^0\end{pmatrix}$ is a $Z_2$ scalar doublet, which we assume does not get a VEV, and $\phi$ is a scalar singlet which acts as the DM candidate responsible for 27\% of the energy density of the universe. The relative masses are such that $\phi$ is lighter than $\rho$ and $L$, and $\rho$ can decay to $\phi$ via a Higgs coupling term given by $\phi^\ast\rho H$. $\bar{L}_v$ is a vector-like heavy $Z_2$ odd lepton doublet, and $\nu_R$ is a singlet right handed neutrino with a mass of $m_{\nu_R} \simeq$ 1 MeV. In such a scenario, the $\phi$ annihilates into a pair of $\nu_R$ via a triangle loop containing $L^{\pm},\,\rho^{\mp}$ or  $L^{0}_v,\,\rho^{0}$. A photon can be emitted from any of the internal charged legs associated with $L^{\pm}_v,\,\rho^{\mp}$ to make the final state $\nu\bar{\nu}\gamma$.  

\section{Dark matter and the 21 cm line} 
\label{sec:21cm}

\par Recently the Experiment to Detect the Global Epoch of Reionization Signature (EDGES) collaboration reported a measurement of the absorption of the redshifted 21 cm line from hydrogen gas at a redshift of $z \approx 15-20$~\cite{Bowman:2018yin}. This result can be interpreted as demonstrating a stronger absorption signal than standard astrophysical expectations, and has sparked a flurry of studies in the dark matter literature, including implications for dark matter-baryon couplings~\cite{Fialkov:2018xre,Berlin:2018sjs,Barkana:2018qrx,Fraser:2018acy,Kang:2018qhi,Slatyer:2018aqg,Hirano:2018alc,Mahdawi:2018euy}, dark matter annihilation~\cite{DAmico:2018sxd,Yang:2018gjd,Cheung:2018vww,Liu:2018uzy}, decaying dark matter~\cite{Clark:2018ghm,Liu:2018uzy,Mitridate:2018iag}, primordial black holes~\cite{Clark:2018ghm,Hektor:2018qqw}, fuzzy dark matter~\cite{Lidz:2018fqo}, dark sectors~\cite{Costa:2018aoy,Pospelov:2018kdh,Li:2018kzs,Jia:2018csj}, and non-cold dark matter models including warm dark matter and axions~\cite{Safarzadeh:2018hhg,Lambiase:2018lhs,Lawson:2018qkc,Moroi:2018vci,Schneider:2018xba,Sikivie:2018tml}.

\par After recombination, neutral hydrogen gas was kept in thermal equilibrium with the cosmic microwave background via Compton scattering from free electrons until a redshift of $z \approx 150$, when it thermally decoupled from the CMB. As a non-relativistic component, the gas temperature, $T_{\rm{G}}$, will subsequently cool with redshift as $(1+z)$ relative to the CMB temperature, $T_{\rm{CMB}}$. The fractional amount of neutral hydrogen, $x_{\rm{H_I}}$ will be close to unity before re-ionization, and will reside predominantly in the 1S ground state, which has a hyperfine splitting into singlet and triplet states due to the interaction of the proton and electron magnetic moments. The energy difference between these states corresponds to a frequency of 1420.4 MHz or a wavelength of 21.1 cm, with the relative population of the triplet state (of number density $n_1$), and singlet state (of number density $n_0$), characterized by the spin-temperature, $T_{\rm{S}}(z)$, through the relation $n_1/n_0 = 3e^{-0.068{\rm{K}}/T_{\rm{S}}}$, where 0.068 K is the equivalent temperature of the 21.1 cm wavelength corresponding to the hyperfine energy splitting $\Delta E = 5.9\times 10^{-6}\,\rm{eV}$. After decoupling, the expected temperature relation is $T_{\rm{G}} < T_{\rm{S}} < T_{\rm{CMB}}$, with $T_{\rm{S}}$ approaching $T_{\rm{CMB}}$ as interactions with CMB photons flip the hyperfine state and draw the spin temperature towards the CMB temperature. After the onset of star formation, Lyman-$\alpha$ photons will recouple $T_{\rm{S}}$ to $T_{\rm{G}}$ via the Wouthuysen-Field effect~\cite{Wouthuysen:1952,Field:1958,Hirata:2005mz,Breysse:2018slj}. 

\par The measurement of the 21 cm line is a target of several current and future observations, and is projected to provide a wealth of new cosmological data that will shed light on the so-called Dark Ages of cosmology prior to star formation (for reviews see, for example,~\cite{Furlanetto:2006jb,Pritchard:2011xb}). The experimental signature of the 21 cm line is determined by the difference between the spin temperature and the CMB temperature through the brightness temperature relation
\bea
T_{21}(z) \approx 0.023{\rm{K}}\left(\frac{0.15(1+z)}{10\Omega_{m}}\right)^{1/2}\left(\frac{\Omega_bh}{0.02}\right)\left(1 - \frac{T_{\rm{CMB}}(z)}{T_{\rm{S}}(z)}\right)x_{\rm{H_I}}(z)
\label{equ:T_21}
\eea
For the the total matter and baryonic fractions of the critical energy density of the Universe, we adopt the values $\Omega_m = 0.3$ and $\Omega_b = 0.04$, respectively, and we use $h = 0.7$ in units of $100\,{\rm{km}}\cdot{\rm{s}}^{-1}\cdot{\rm{Mpc}}^{-1}$ for the Hubble parameter.

\par The EDGES measured an absorption signature of $T_{21} = -500^{+200}_{-500}$ mK (99\% C.L.) at a redshift of $z \approx 17.2$ with a central value of 78 MHz. This measurement implies a gas temperature that is about a factor of two lower than that expected by conventional astrophysical modeling. A lowered gas temperature can be used to constrain interactions that can lead to a cooling effect, but it can also be used to constrain any processes that would lead to a heating of the gas temperature through injection of energy, thereby reducing any absorption signal.

\par The evolution of the electron fraction and the hydrogen gas temperature will be altered in the presence of energy injection from dark matter annihilation. Annihilations will increase $x_e$ relative to the standard expectations, and these additional electrons can alter the initial CMB-baryon decoupling, as well as injecting energy into the hydrogen gas, with a subsequent rise in $T_{\rm{G}}$. The energy injection depends on the fraction of dark matter that is annihilating, $f_{\rm{DM}}$, the dark matter energy density, $\rho_{\rm{DM}}$, the annihilation cross section $\langle\sigma v\rangle_{\rm{ann}}$, and dark matter mass, $m_{\rm DM}$, through the relation
\bea
\frac{{\rm{d}}E}{{\rm{d}}V{\rm{d}}t} = \rho_{\rm{DM}}^2f_{{\rm{DM}}}^2\frac{\langle\sigma v\rangle_{\rm{ann}}}{m_{\rm DM}}
\eea
Now that we have the dependence on the annihilations for the relevant observables, we can employ the EDGES result to constrain the dark matter annihilation models described above.


\section{Results} 
\label{sct:results} 

In this section we use Fermi-LAT diffuse gamma-ray data and and data from dSphs, as well as the 21 cm observation of EDGES, to constrain the primarily two to two $p$-wave component DM annihilation models via their $s$-wave components arising from radiating a photon or electroweak gauge boson. 

\subsection{Gamma-ray constraints}
\label{sec:results_fermi}  

\subsubsection{Diffuse gamma-ray data} 
We begin discussing the constraints from diffuse gamma-ray data.  Our data selection and analysis method generally follow that of Refs.~\cite{Bringmann:2012vr,Ackermann:2015lka}; we will note the particular aspects in which they differ. We use Fermi Science Tools version v11r5p3~\footnote{\url{https://fermi.gsfc.nasa.gov/ssc/data/analysis/software/}}, and select Pass 8 SOURCE-class events for mission elapsed time 239557417 s to 554861541 s. We apply the recommended  \texttt{(DATA\char`_QUAL>0)\&\&(LAT\char`_CONFIG==1)} filter to ensure quality data and a zenith cut $z_{max} = 100^\circ$ to filter background gamma-ray contamination from the Earth's limb. 

\par For our Region-of-Interest (ROI), we take the R90 region as defined in Ref.~\cite{Ackermann:2015lka}, which corresponds to a cut on the photon direction of 90$^\circ$ from the Galactic center. This amounts to taking data from half of the sky, with the regions in the Galactic plane, corresponding to longitudes $>6^\circ$ and Galactic latitudes $>5^\circ$, masked out. In this region, we take the dark matter density profile to be isothermal, $\rho(r) = \rho_0/[ 1+(r/r_s)^2]$, where $r_s=5$ kpc, and $\rho_0$ is normalized so that the dark matter density at the location of Sun is $\rho \left(r_{\odot} = 8.5 \ \rm{kpc} \right)=0.4 \ \rm{GeV \ cm^{-3}}$. This density profile was chosen because it provides the least stringent constraints on the models that we consider. We have verified this by examining the impact of alternative distributions, in particular NFWc from Ref.~\cite{Ackermann:2015lka}. For the NFWc profile constraints, we use photons within only a $3^\circ$ angle from the Galactic center.




\par To produce constraints for a given dark matter mass, $m_{\rm DM}$, we consider photons within the energy range $0.4 m_{\rm DM} < E_\gamma < 2.25 m_{\rm DM}$. The lower bound is set in order to contain photons from the peak of the bremsstrahlung emission of the spectrum. The maximum energy was chosen so as to include a large enough sample of background photons, to ensure that the background is well fit by a power law. We have verified that for our entire mass range, we are in the regime in which our uncertainties are dominated by statistics rather than systematics, so that the power law fit for the background is a good description of the data. Within this energy range we perform a binned likelihood analysis, with photons in equal spaced logarithmic bins, with 50 bins per decade. We determine the best fit power law index for the background, and then generate a new set of pseudo-data from this fit. We then fit this psuedo-data to a model which is a sum of a background plus the line from the bremsstrahlung peak. For a given model, we define the limits as where the TS statistic for the Log-Likelihood exceeds TS $ > 1.355$. 

\begin{figure*}[ht]
\begin{tabular}{cc}
\includegraphics[height=5cm]{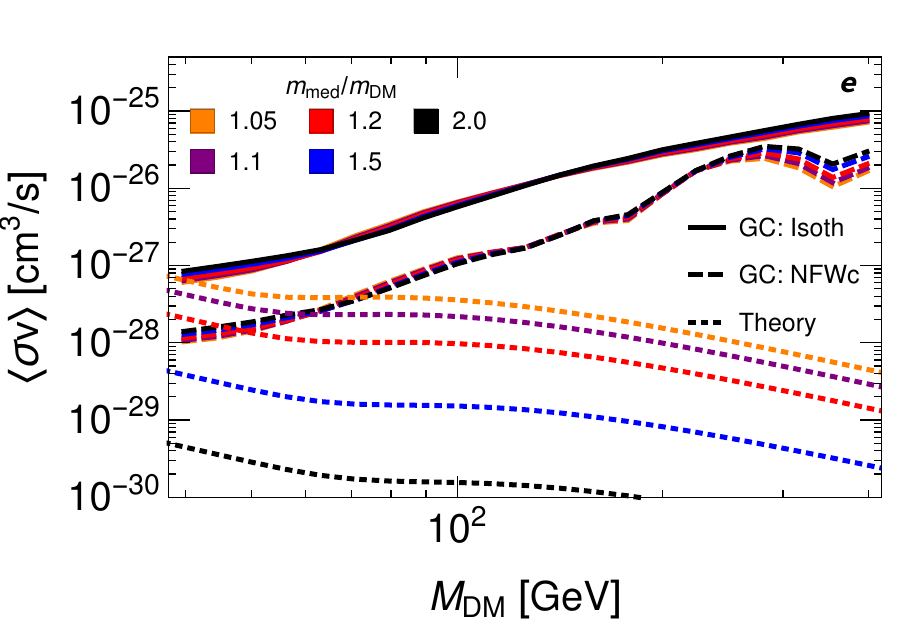} &
\includegraphics[height=5cm]{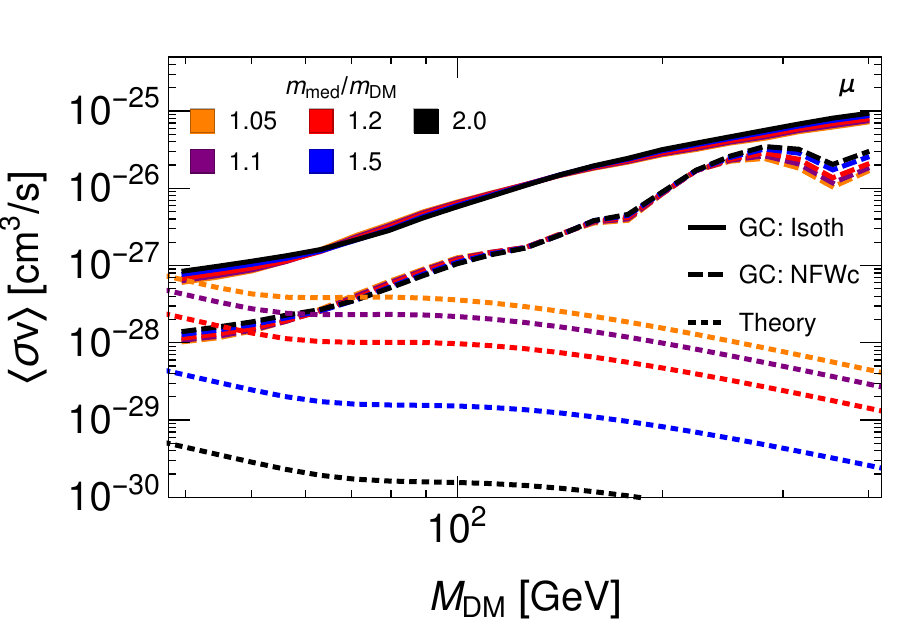}\\
\end{tabular}
\begin{tabular}{cc}
\includegraphics[height=5cm]{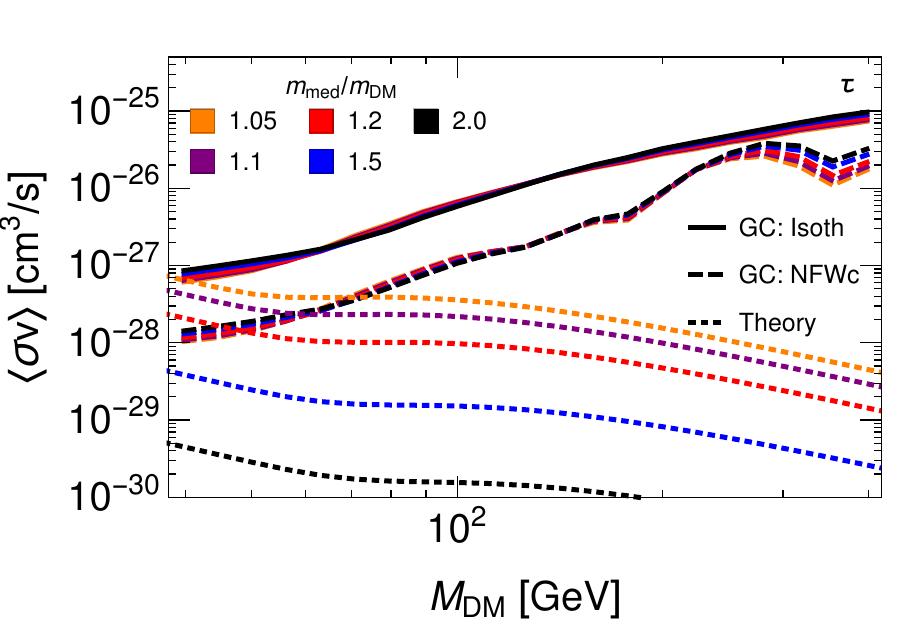} &
\includegraphics[height=5cm]{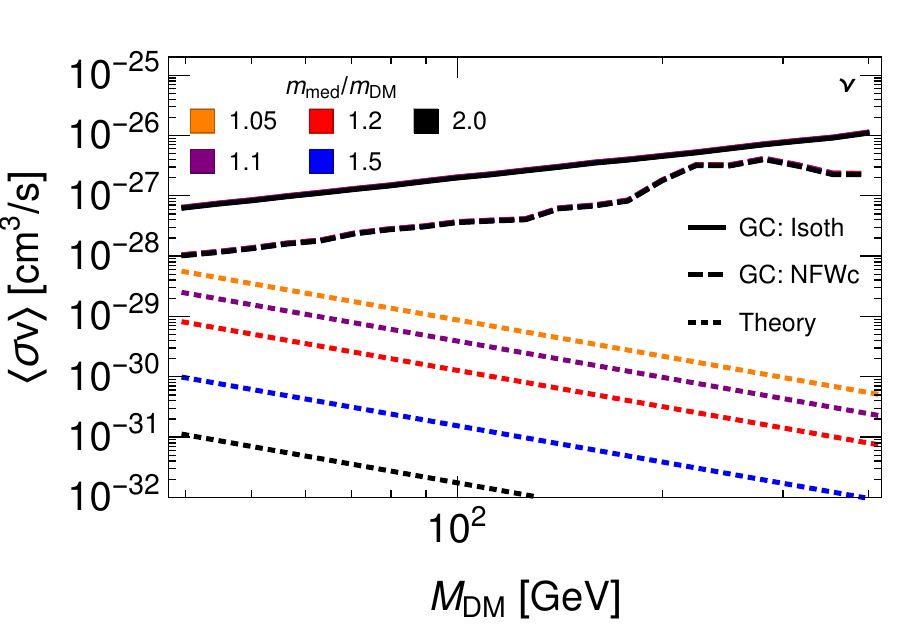}\\
\end{tabular}
\caption{Constraints on the annihilation cross section using null detections of gamma-ray lines from Fermi-LAT data (solid lines). These constraints assume the isothermal profile, as defined in the text. The theoretically-calculated cross section for various mediator mass ratios, $m_{\rm med}/m_{\rm DM}$, are shown as dashed lines. Lepton final states are electron (top left), muon (top right), tau (bottom left), and neutrino (bottom right). The gamma-ray lines constraints using the NFWc dark matter profile are shown as short dashed.}
\label{fig:GC_comb_th}
\end{figure*}

\par The resulting constraints from gamma-ray lines are shown in Fig.~\ref{fig:GC_comb_th} for $f\bar{f}+(\gamma, W, Z)$ final states. We also show the thermally-averaged cross sections for each of these final states for the scenarios where the DM primarily annihilates into $f\bar{f}$ ($p$-wave dominated). Various colored lines, both for experimental and theory scenarios, are shown for different mediator to DM mass ratios. The $e,\,\mu,\,\tau$ final states shown  are based on the MSSM model where the neutralino and slepton mass differences vary between 5\% and 100\%. We note that though we use SUSY for the purposes of an example, this analysis can be applied to any classes of models including $t$- and $u$-channel scalar mediators with similar mass ratios to that of the DM. LHC searches for slepton masses leave a large amount of unconstrained parameter space for the selectron and smuon for mass differences of $\Delta M(m_{{\tilde e},\, {\tilde\mu},\,\tilde\chi^0_1})\leq 60$ GeV with respect to the neutralino DM particle~\cite{Aaboud:2017leg}, and stau masses are constrained to be $m_{\tilde{\tau}}>100$ GeV from LEP limits~\cite{Heister:2001nk,Abdallah:2003xe,Achard:2003ge,Abbiendi:2003ji} (LHC limits on stau masses are approaching a similar level~\cite{Sirunyan:2018vig}). 


\par The constraints in Figure~\ref{fig:GC_comb_th} can be used to place constraints on various dark matter scenarios.  
We find that the constraint rules out a DM mass $m_{\rm DM} \lesssim 30$ GeV for a mediator to DM mass ratio of $\simeq 5\%$ with the NFWc profile while the constraint becomes 70 GeV for e, $\mu$ and $\tau$ final states. We used the MSSM parameter space for the charged lepton scenarios. For the $\nu\nu\gamma$ final state we use the model described above with $\lambda\sim 2$ and we assume all the charged heavy state masses are the same.

\begin{figure*}[ht]
\begin{tabular}{cc}
\includegraphics[height=5cm]{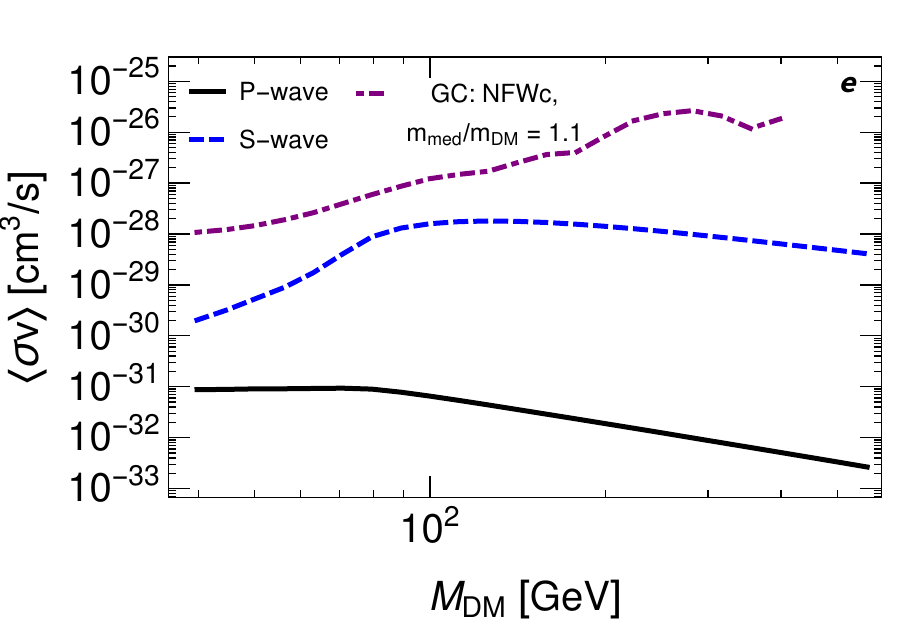} &
\includegraphics[height=5cm]{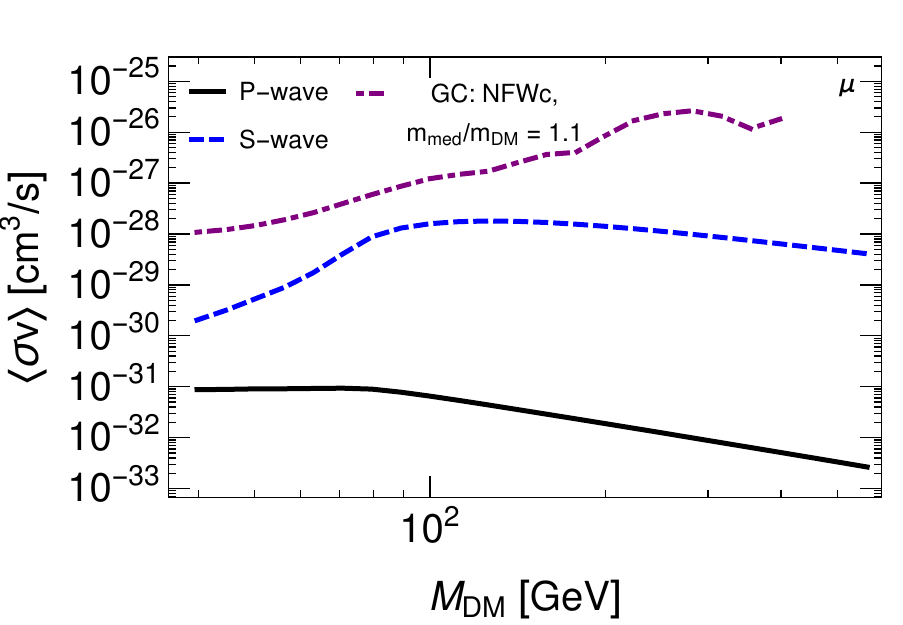}\\
\end{tabular}
\begin{tabular}{cc}
\includegraphics[height=5cm]{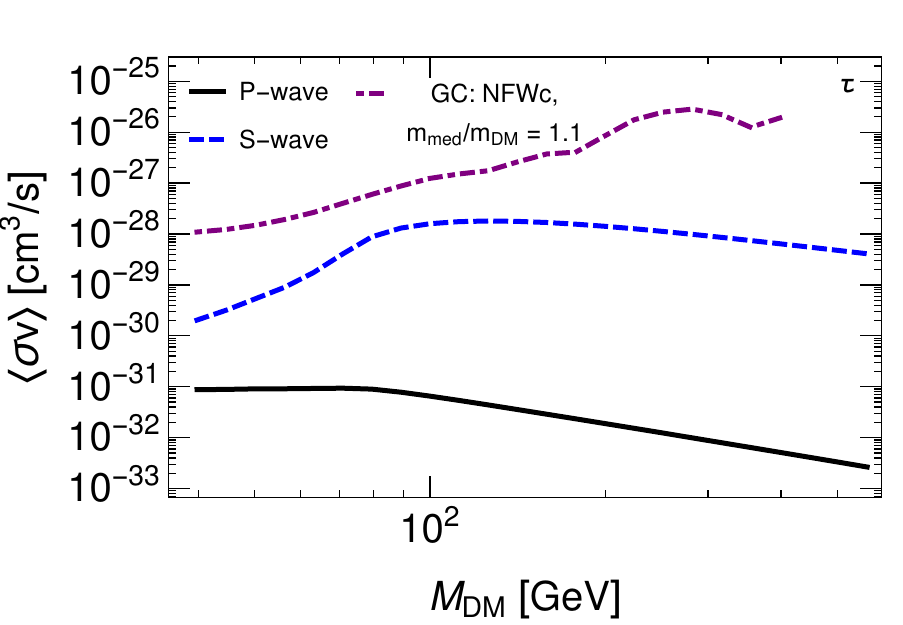} &\\
\end{tabular}
\caption{Upper bound constraints imposed by the Galactic center (long-short dashed) and the theoretical calculated cross section for various final states which satisfy the thermal dark matter abundance. The dashed lines show the s-wave and the solid lines show the p-wave component today. }
\label{fig:GC_comb_th_st}
\end{figure*}

In Figure~\ref{fig:GC_comb_th_st}, we show the current $s$- and $p$-wave annihilation rates required to produce the appropriate DM abundance. Here we have calculated the mediator to DM mass ratios necessary for obtaining the relic abundance with micrOMEGAs~\cite{Belanger:2001fz, Belanger:2004yn}. In comparison with Figure~\ref{fig:GC_comb_th}, the thermal DM line corresponds closely with $m_{\rm med}/m_{\rm DM}=1.05$ for DM masses above 100 GeV. The larger mass ratios presented in the figure can arise in non-thermal scenarios described below. We find that the $p$-wave component today is small except in the case of lighter DM masses. When the DM mass becomes small, the mass difference between the mediator and DM increases. As the mass ratio increases, the $s$-wave component is suppressed. We see that the current reach is nearly an order of magnitude from the $s$-wave component for a DM mass around 100 GeV. We do not show the neutrino final state figure here because the $p$-wave component is model dependent, e.g., based on our Lagrangian we do not have any $\bar{\nu}\nu$ final states unless we assume large values of Dirac mass for the neutrinos. In such a scenario, the $\nu\nu\gamma$ final state would provide the necessary relic abundance.

The non-thermal scenario can be constrained from Figure~\ref{fig:GC_comb_th}. The non-thermal picture emerges generically in UV theories like string theory due to the presence of gravitationally coupled scalars~\cite{Coughlan:1983ci, Banks:1993en, deCarlos:1993wie, Acharya:2008bk, Dutta:2009uf, Allahverdi:2013noa} which are  displaced from their minimum during inflation which can be of order $M_P$~\cite{Dine:1995kz}. After the end of inflation, when $H\leq m_{\rm mod}$, the moduli start dominating the energy density of the universe which gets reheated when the moduli decay.
Since the moduli are  gravitationally coupled, they tend to decay very late with
a reheating temperature 
$T_{\rm{rh}} \sim \sqrt{\Gamma M_P}\sim m_{\rm mod} \sqrt{\frac{m_{\rm mod}}{M_P}}$, where $\Gamma$ is the decay width of the modulus and $m_{\rm mod}$ is the mass of the moduli. $T_{\rm{rh}}$
needs  to be larger than $T_{\rm BBN}$ in order to maintain the successful BBN predictions.

If we use the NFWc profile, we can constrain the reheat temperature ($T_{\rm{rh}}>0.5$ GeV) for a DM mass $\sim$ 100 GeV (with the freeze-out temperature, $T_{\rm{f}}\sim$ 5 GeV) for charged lepton final states for MSSM parameter space. The constraint on $T_{\rm{rh}}$ is model dependent since the DM annihilation calculation not only depends on the mass scales of DM, mediator, and final states but also on the couplings. In the case of the MSSM, the coupling is $g_{\rm{ weak}}$ which can be different for other models leading to larger annihilation cross sections and a decrease of the lower limit on $T_{\rm{rh}}$. 

In a non-thermal  scenario, the dark matter abundance is given by the following expression~\cite{Moroi:1999zb, Aparicio:2016qqb}
\begin{equation}
\label{equ:NonThermalAbundance}
\left(\frac{n_\chi}{s}\right) = {\rm{min}} \left\{\left(\frac{n_\chi}{s}\right)^{\rm obs} \, \frac{\langle \sigma v\rangle^{\rm th}}{\langle \sigma v\rangle} \, \sqrt{\frac{g_*(T_{\rm f})}{g_*(T_{\rm rh})}} \, \frac{T_{\rm f}}{T_{\rm rh}}, \quad Y_{\phi} \rm{Br}_{\phi}\right\} \,,
\end{equation}
$\left(\frac{n_\chi}{s}\right)^{\rm obs} \simeq \Omega^{\rm obs} \, \left(\frac{\rho_{\rm crit}}{m_{\chi s h^2}}\right)$, while $Y_{\phi} \simeq \frac{3 T_{\rm rh}}{4 m_\phi}$ is the yield of DM abundance from modulus decay, and Br$_{\phi}$ is the branching ratio of the modulus decay into R-parity odd particles. The first term refers to the {Annihilation Scenario}, while the second term refers to the {Branching Scenario}. In the Branching Scenario, the dark matter is frozen-in and the value of the cross section is only bounded from above.  In Fig.~\ref{fig:GC_comb_th}, the annihilation scenarios are constrained.

\subsubsection{Dwarf spheroidals}

We now move on to discuss constraints using gamma-ray data from dSphs. In this case, rather than using the photon data directly, we use the pre-generated likelihood functions provided in Ref.~\cite{Fermi-LAT:2016uux}. The photon flux observed at Earth is calculated through the relation
\begin{equation}
\Phi = \frac{1}{8\pi}\frac{\langle \sigma v \rangle}{m^2_{\rm DM}} \times J,
\label{equ:flux}
\end{equation}
where $J$ is the $J$-factor, which incorporates the DM distribution within the dSph as well as its distance from the observer. For our analysis we adopt the $J$-factor values used in Ref.~\cite{Ackermann:2015zua}. A log-likelihood analysis is then performed on the combined system of all the dSphs to obtain the null likelihood probability for the flux amount~\cite{Ackermann:2015zua}.

\begin{figure*}[ht]
\begin{tabular}{cc}
\includegraphics[height=5cm]{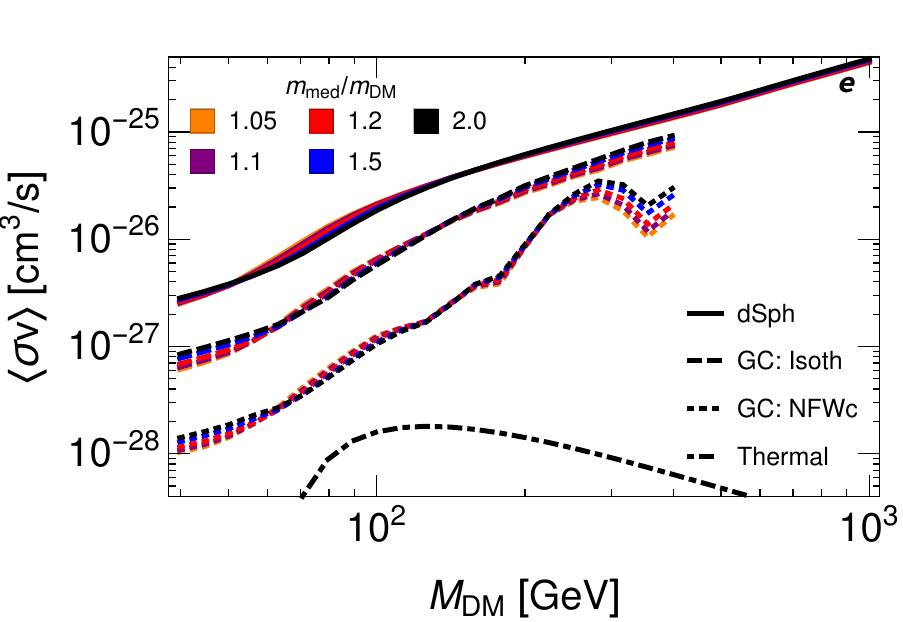} &
\includegraphics[height=5cm]{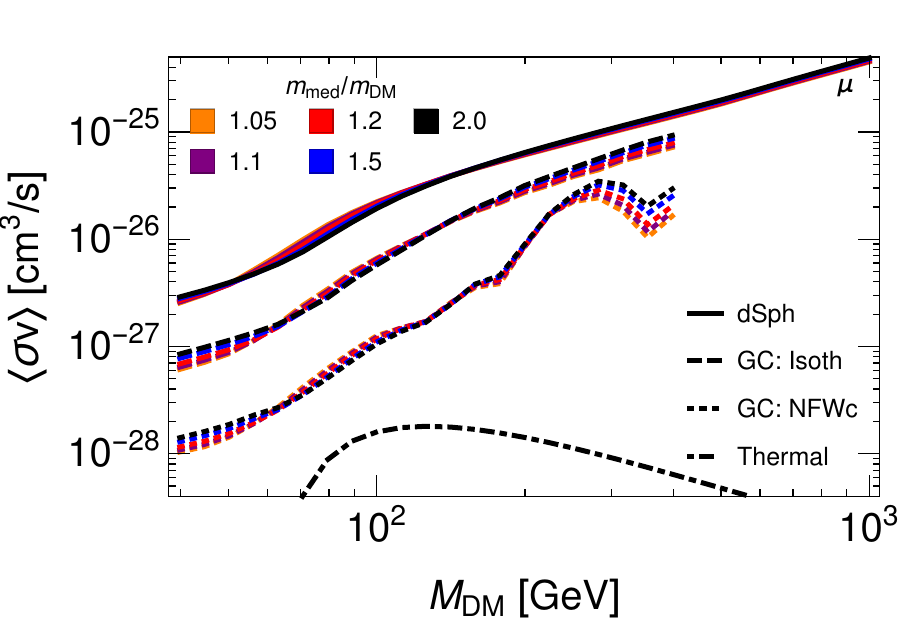}\\
\end{tabular}
\begin{tabular}{cc}
\includegraphics[height=5cm]{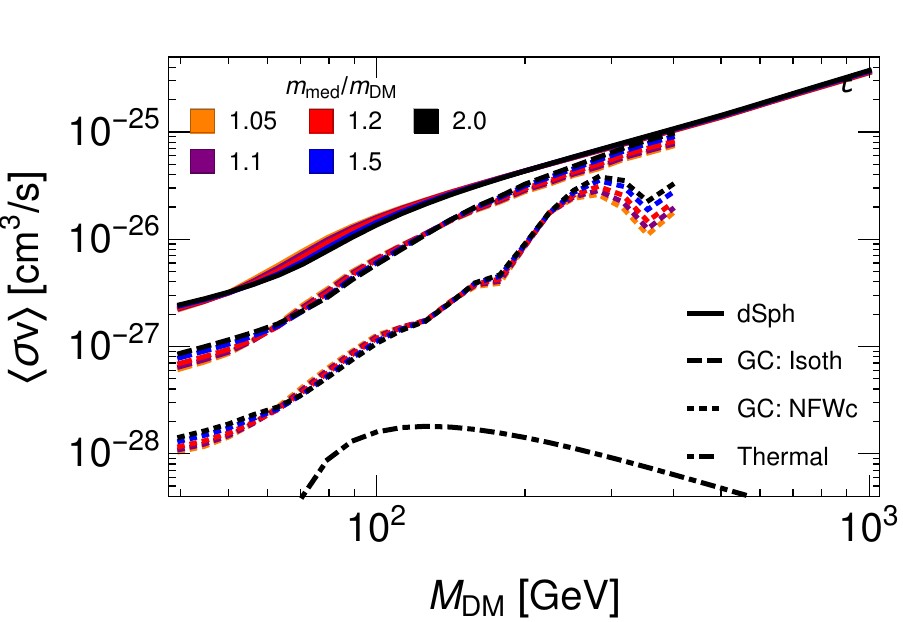} &
\includegraphics[height=5cm]{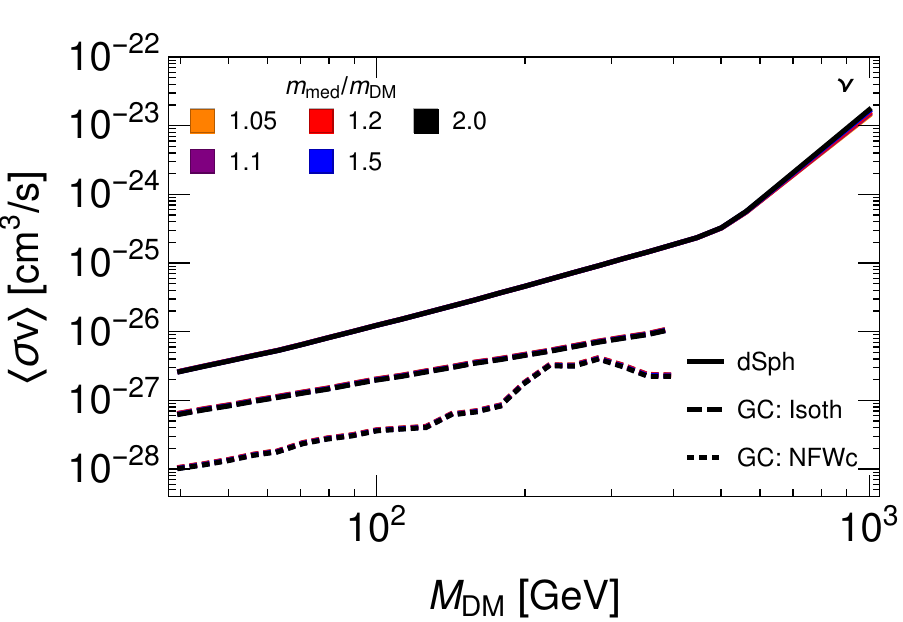}\\
\end{tabular}
\caption{Upper bound constraints imposed by dSph (solid), the Galactic center (dashed) and the theoretical calculated cross section for thermal dark matter (long-short dashed) for various mediator mass ratios, $m_{\rm med}/m_{\rm DM}$. Lepton final states are electron (top left), muon (top right), tau (bottom left), and neutrino (bottom right). We also show the Galactic center using a less conservative dark matter profile (short dashed).}
\label{fig:GC_comb_dsph}
\end{figure*}

In Figure~\ref{fig:GC_comb_dsph}, we compare the dSph and diffuse gamma-ray constraints for $e^+e^-$, $\mu^+\mu^-$, $\tau^+\tau^-$, and $\nu\nu$ final states in conjunction with a final state photon. Generally across the entire mass range, we find that the constraints from diffuse gamma-ray data are more stringent than those from dSphs.

The constraints we find can be compared to those previous found in Ref.~\cite{Bringmann:2012vr}, which uses an approach similar to ours. At $m_{\rm DM}\sim90$ GeV, there is a slope change that is present in all of our cases. This is a direct result of the introduction of $W/Z$ boson channels becoming dominant pathways. The remaining differences between our results and Ref.~\cite{Bringmann:2012vr} may be attributed to the binning resolution for the data and the model.

\subsection{Constraints arising from 21 cm signal}
\label{sec:results_21cm} 

\par Our approach in calculating constraints on the 21 cm follows the approach used in Ref.~\cite{Clark:2018ghm}. We developed an effective efficiency map for the specific annihilation spectra of each model utilizing the electron and photon effective efficiencies from Ref.~\cite{Slatyer:2012yq,Liu:2016cnk,Clark:2016nst}. These effective efficiencies capture the particle interaction with the medium and characterize how much energy is deposited from the annihilation into causing changes to the hydrogen ionization fraction and gas temperature. The effective efficiencies are then incorporated in calculating the ionization and temperature history of the Universe during the periods of recombination up to reionization through our modified version of Hyrec~\cite{AliHaimoud:2010dx}. From this history, $T_{\rm S}$ in Eq.~\ref{equ:T_21} can be calculated through~\cite{Zaldarriaga:2003du}
\bea
T_{\rm S} & = & \frac{T_{\rm{CMB}}+y_{\rm c} T_{\rm G}+y_{\rm{Ly\alpha}}T_{\rm{Ly\alpha}}}{1+y_{\rm c}+y_{\rm{Ly\alpha}}}, \label{eq:T_s}\\
y_{\rm c} & = & \frac{C_{10}}{A_{10}} \frac{T_\star}{T_{\rm G}}, \label{eq:y_c}\\
y_{\rm{Ly\alpha}} & = & \frac{P_{10}}{A_{10}} \frac{T_\star}{T_{\rm{Ly\alpha}}}, \label{eq:y_Lya}
\eea
where $C_{10}$ is the collisional de-excitation rate of the triplet hyperfine level, $A_{10}=2.85\times10^{-15}$s$^{-1}$ is the transition's spontaneous emission coefficient, $T_\star=h\nu_0/k_{\rm B}=0.068$ K is the temperature of the Lyman-Alpha photon, and $P_{10}\approx 1.3\times 10^{-12} S_\alpha J_{-21}$s$^{-1}$ is the indirect de-excitation rate due to Lyman-Alpha photon absorption with $S_\alpha$ being a factor of order unity that incorporates spectral distortions~\cite{Hirata:2005mz}. $J_{-21}$ is the Lyman-Alpha background intensity in units of $10^{-21}$ erg cm$^{-21}$ s$^{-1}$ Hz$^{-1}$ sr$^{-1}$ For $J_{-21}$, we use the results from Ref.~\cite{Ciardi:2003hg} as an example of the strong coupling limit which produces the weakest constraints. Finally the annihilation rate is adjusted to defined limits (ie. $-100$ mK) to place conceptual constraints on the impact of the 21 cm signal.

\begin{figure*}[ht]
\begin{tabular}{cc}
\includegraphics[height=5cm]{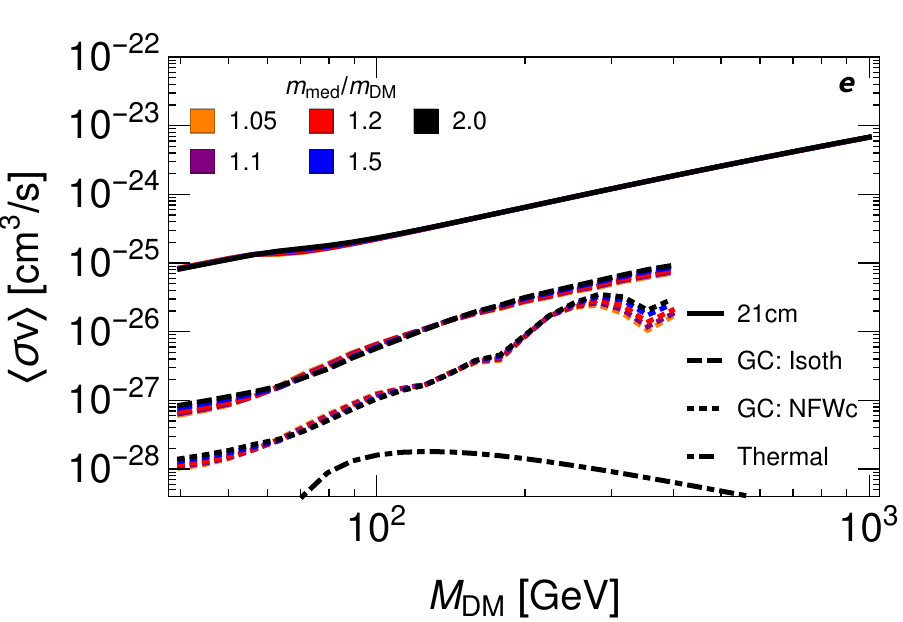} &
\includegraphics[height=5cm]{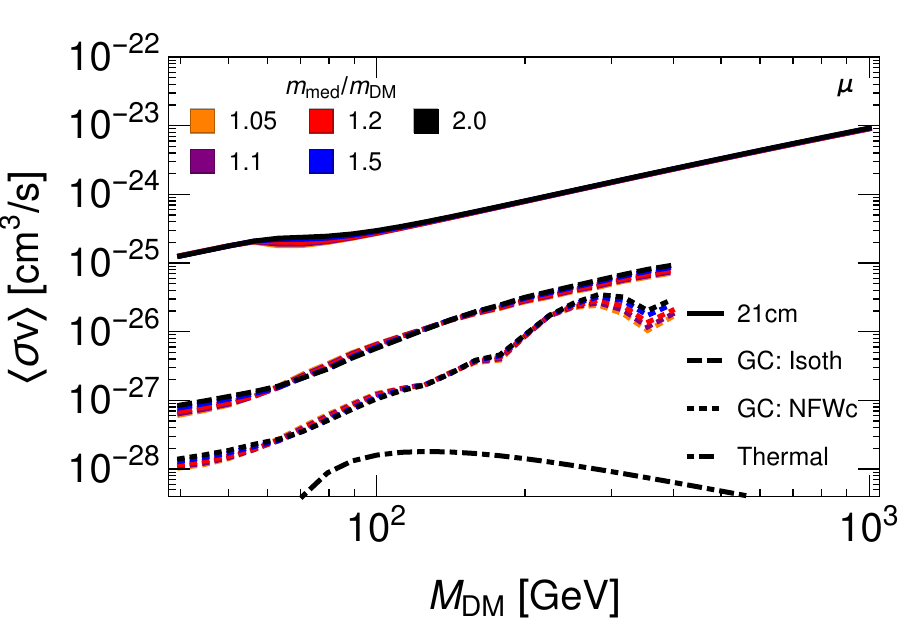}\\
\end{tabular}
\begin{tabular}{cc}
\includegraphics[height=5cm]{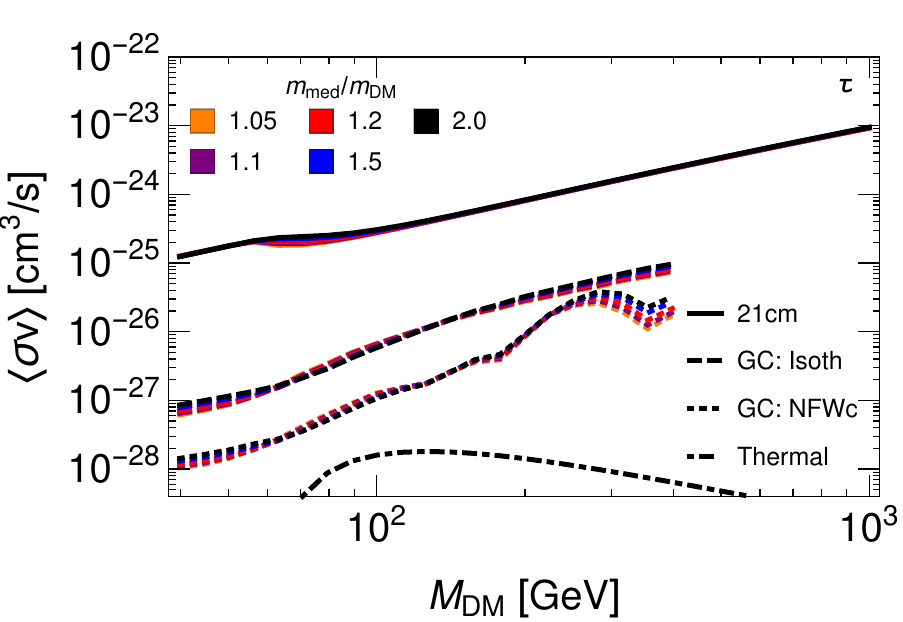} &
\includegraphics[height=5cm]{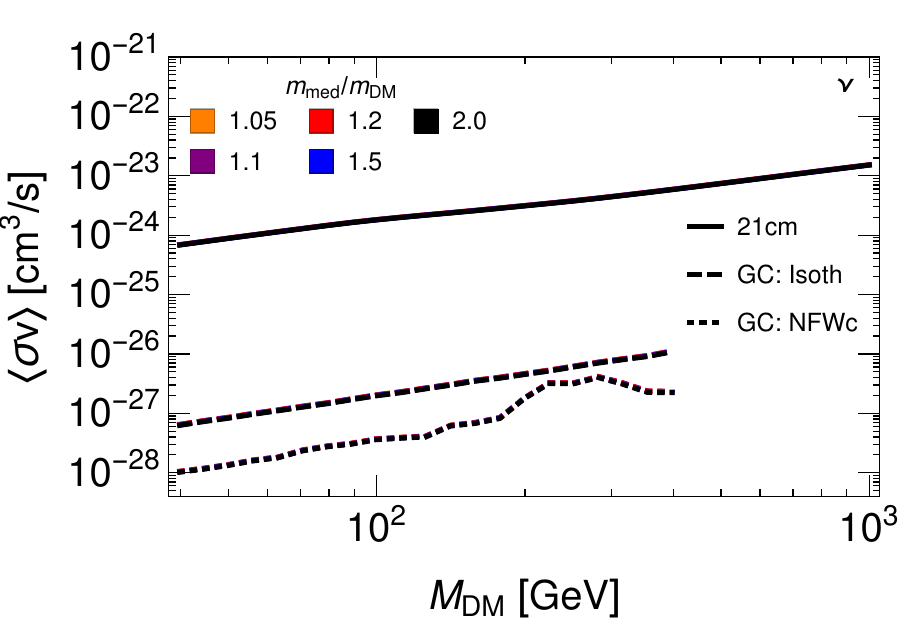}\\
\end{tabular}
\caption{Upper bound constraints imposed by 21 cm with $\Delta T = -100$ mK (solid), the Galactic center (dashed) and the theoretical calculated cross section for thermal dark matter (long-short dashed) for various mediator mass ratios, $m_{\rm med}/m_{\rm DM}$. Lepton final states are electron (top left), muon (top right), tau (bottom left), and neutrino (bottom right). We also show the Galactic center using a less conservative dark matter profile (short dashed).}
\label{fig:GC_comb_21cm}
\end{figure*}

\par By requiring the $T_{21}$ correction relative to its standard astrophysical value at $z\sim 17$ to be less than 100 and 150 mK (namely $T_{21}(z=17)<-100$ and $-50$ mK respectively), we show the constraints for the $p$-wave dominated model in Fig.~\ref{fig:GC_comb_21cm}. We also compare the 21 cm and diffuse constraints for $f\bar{f}+\gamma/W/Z$ and find that the galactic center constraint from Fermi are currently much more constraining than 21 cm observations. In contrast to the diffuse constraints, the $\bar{\nu}\nu\gamma$ final state is constrained at about the same level as the other leptonic final states, rather than providing the most stringent bounds. Although there are currently a few orders of magnitude separating the 21 cm and diffuse bounds, there are upcoming 21 cm observations which can increase the competitiveness of this bound, thus providing a useful tool from a different cosmic epoch in the investigation of $p$-wave models. The Cosmic Microwave Background (CMB) is an additional method to be considered. Figure~\ref{fig:21cm_CMB_compare} compares the 21 cm results to those of the CMB. The current 21 cm results are comparable to the CMB with the CMB being slightly stronger. However, they are both much weaker than both the dSph and diffuse constraints.



\section{Summary}
\label{sec:summary}

\par If dark matter self-annihilates to Standard Model final states, then its annihilation cross section is a fundamental property which current experiments can explore. The dominant annihilation channel can vary over the course of cosmic history due to a velocity dependence that will dramatically suppress $p$-wave annihilations relative to $s$-wave for non-relativistic dark matter. If dark matter were to annihilate dominantly through a $p$-wave process, then it will be observationally very challenging to probe such models with standard indirect detection techniques. In this work we have investigated well-motivated models which annihilate dominantly via a $p$-wave process to two-body final states in the very early universe, but can have a leading three body final state annihilation when the dark matter is non-relativistic. The cross section is enhanced by the well known mechanism of internal and final state vector boson bremsstrahlung of $W/Z/\gamma$, leading to $f\bar{f}+W/Z/\gamma$ final states. 

\par The model framework we have adopted is a rather general, SUSY-inspired model with Majorana dark matter of mass $m_{\rm DM}$ annihilating via $t$- and $u$-channel exchange of charged mediators of mass $m_{\rm med}$. Annihilation to three-body final states is enhanced as the mediator mass approaches that of the dark matter, and we have included the dependence on this ratio in our analysis.  We find that the bounds are fairly insensitive to $m_{\rm med}/m_{\rm DM}$ as it is varied from 1.05 to 2.0. We have employed complementary aspects of the different final state bosons in order to strengthen the bounds on dark matter annihilation. Specifically, the photon bremsstrahlung can produce line-like features which can be constrained with data from the Fermi satellite via well-known line search techniques. As the dark matter mass is increased, the parameter space to produce the massive $W$ and $Z$ bosons becomes available, providing complementary signals to the photon line search through the addition to the continuum spectrum produced by the $W$ and $Z$ decays.

\par Some aspects of this work to highlight are the use of recent 21 cm observations to constrain annihilations with vector bremsstrahlung, and the development of constraints using the $\bar{\nu}\nu\gamma$ final state. Developing bounds on dark matter physics from 21 cm observations are quickly becoming a standard tool in the field, though the limits derived in the current work are significantly weaker than those from dSph and diffuse data from the Fermi satellite searches. Final states consisting of neutrinos accompanied by no other particles, or dominantly annihilating to neutrinos without the existence of charged lepton final state channels, lead to a very difficult search. However, we have demonstrated that in some dark matter models, $\bar{\nu}\nu\gamma$ final states can actually provide leading constraints compared to those from charged leptons, $\ell^+\ell^-\gamma$, from diffuse and dSph searches, with the 21 cm observations for $\bar{\nu}\nu\gamma$ final states producing bounds competitive with those from $\ell^+\ell^-\gamma$.

Although models of dark matter dominantly annihilating to two-body final states through $p$-wave processes are quite challenging to probe observationally, we see that the situation is not hopeless. 
We found that some DM masses are constrained while for non-thermal scenarios the reheating temperature $T_{\rm{rh}}$ gets constrained. 
By investigating scenarios where three-body final states open $s$-wave channels, $p$-wave models can still provide a fertile ground for current and future investigations.

\section{Acknowledgements}
SJC, BD, and LES acknowledge support from DOE Grant de-sc0010813.
JBD thanks Thomas Jacques for clarifying remarks regarding the lepton spectrum from $W/Z$ bremsstrahlung, and the Mitchell Institute for Fundamental Physics and Astronomy at Texas A\&M University for their generous hospitality. JBD acknowledges support from the National Science Foundation under Grant No. NSF PHY-1820801.

\bibliographystyle{utphys}
\bibliography{main}
\newpage
\section*{Appendix}
\begin{figure*}[ht]
\includegraphics[height=5cm]{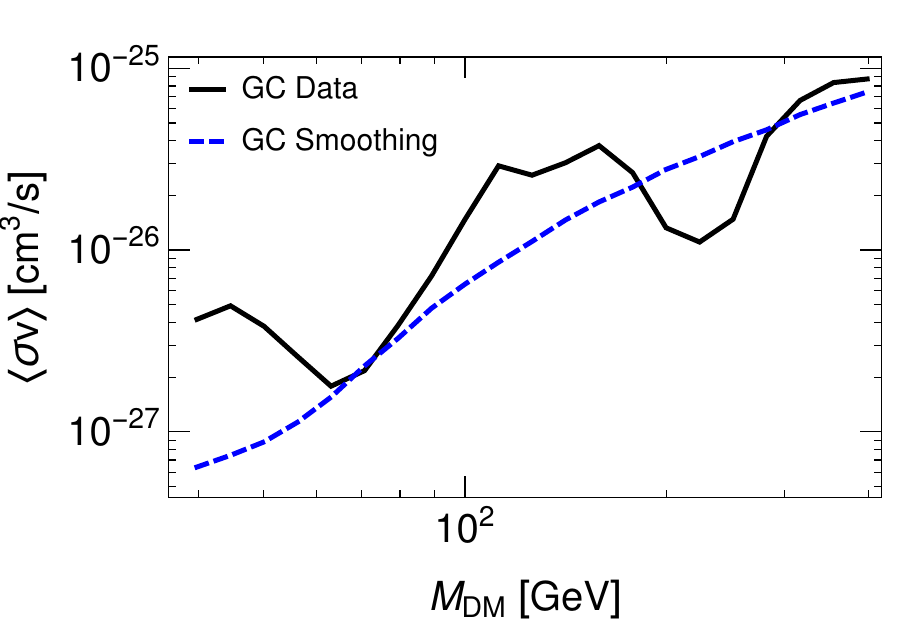}
\caption{Comparison between the results directly obtained from the Galactic center data (solid) and our averaging routine (dashed). The data set used is annihilation to electrons plus a boson with mediator mass ratio 1.1.}
\label{fig:GC_smooth_nonsmooth_compare}
\end{figure*}

In Fig.~\ref{fig:GC_smooth_nonsmooth_compare}, we demonstrate differences that arise between our smoothing algorithm and our analysis that calculates the galactic center constraints straight from the Fermi data. Our smoothing approach fits the data to a power law and then performs the least likelihood analysis off the fit. This approach helps remove statistical fluctuations present in the data. However, we lose the capacity to identify a positive source signal. Comparing the two, the smoothed result is approximately the median result of the baseline result.

\begin{figure*}[ht]
\includegraphics[height=5cm]{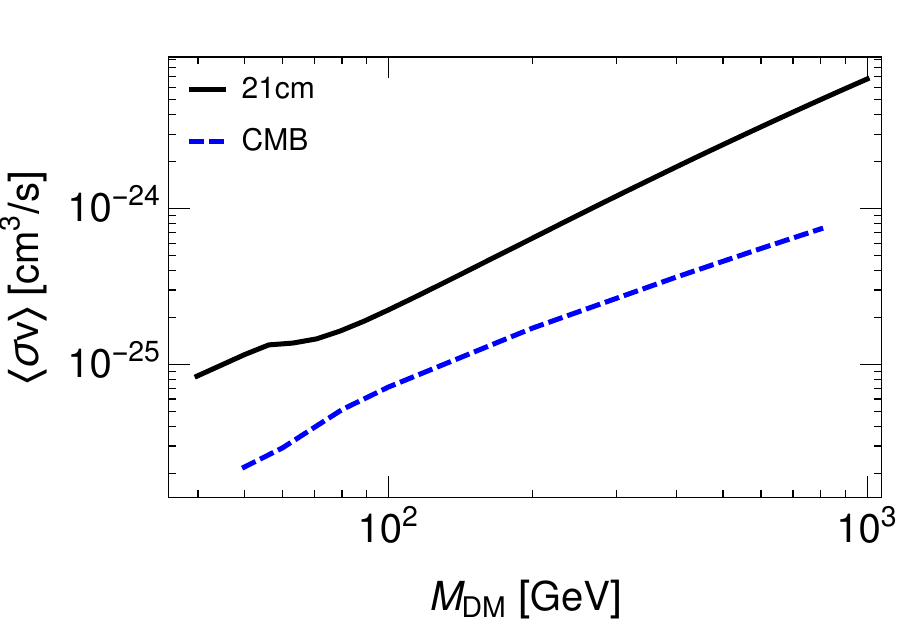}
\caption{Comparison between the results for 21 cm (solid) and CMB (dashed). While the CMB is more constraining than the 21 cm, it is still much weaker than both GC and dSph. The data set used is annihilation to electrons plus a boson with mediator mass ratio 1.1.}
\label{fig:21cm_CMB_compare}
\end{figure*}

The Cosmic Microwave Background (CMB) is another signal we can use to constrain the model. The approach we used to calculate the constraints placed by the CMB is similar to the 21 cm~\cite{Clark:2016nst, Clark:2018ghm}. Fig.~\ref{fig:21cm_CMB_compare} shows the differences between constraints by the CMB and 21 cm. The CMB is more constraining for these models by approximately a factor of 5. However, these bounds are still many orders of magnitude weaker than those set by the Galactic center and dSph.

\end{document}